\documentclass[aps,prl,amsmath,amssymb,twocolumn,nofootinbib,floatfix]{revtex4}
\usepackage{latexsym,amssymb,graphicx,dcolumn,bm,natbib}
\usepackage{epsfig}
\usepackage{subfigure}
\usepackage{amsmath}
\date{\today}
\begin{document}
\bibliographystyle{prsty}

\title{Entropic origin of stress correlations in granular materials}
\author{ G. Lois$^{1,2}$, J. Zhang$^{3}$, T. S. Majmudar$^3$, S. Henkes$^4$, B. Chakraborty$^4$, C. S. O'Hern$^{1,2}$,  and R. P.
Behringer$^3$\\
\normalsize{$^1$Department of Mechanical Engineering, Yale University, New Haven, Connecticut 06520-8284\\
$^2$Department of Physics, Yale University, New Haven, Connecticut 06520-8120, USA\\
$^3$Department of Physics, Duke University, Box 90305,
Durham, NC 27708, USA} \\
$^4$Department of Physics, Brandeis University,
Waltham, MA 02454, USA}
%%% PRL, block 2

\begin{abstract}
We study the response of granular materials to external stress using
experiment, simulation, and theory.  We derive an entropic,
Ginzburg-Landau functional that enforces mechanical stability and
positivity of contact forces. In this framework, the elastic moduli
depend only on the applied stress.  A combination of this feature and
the positivity constraint leads to stress correlations whose shape and
magnitude are extremely sensitive to the applied stress.  The
predictions from the theory describe the stress correlations for both
simulations and experiments semiquantitatively.
\end{abstract}

%%% PRL, block 3
%\pacs{83.80.Fg,45.70.-n,64.60.-i}
%45.70.-n Granular systems
%83.80.Fg Granular solids
%64.60.-i General studies of phase transitions
\maketitle
%%% PRL, block 3

A striking feature of dry granular materials and other athermal
systems is that they form force chain networks in response to applied
stress, such that large forces are distributed inhomogeneously into
linear chain-like structures~\cite{nagelscience,Dinsmore}.  A number
of experimental studies have visualized and quantified these networks
in granular systems using carbon paper \cite{nagel} and photoelastic
techniques~\cite{dantu,jgeng}.  These studies demonstrated that
geometrical and mechanical properties of force chain networks are
acutely sensitive to preparation procedures, especially near the
jamming transition\cite{RBTrush}. For example, in isotropically
compressed systems, force networks are ramified with only short-ranged
spatial correlations of the stress.  In contrast, in sheared systems,
aligned force chains give rise to longer ranged stress correlations
in the compressive direction.  Since granular (or other)
systems near jamming are fragile and highly sensitive to
preparation, one expects that their mechanical properties near jamming
might not be captured by simple linear elastic response~\cite{Landau}.

Developing alternative theoretical descriptions for granular media
is challenging for several important
reasons~\cite{Bouchaud,Cates,Blumenfeld,GoldGold}: (1) since tensile
stresses are absent in dry granular materials they only remain intact
via applied stress, making the limiting zero-stress isostatic state,
where the number of degrees of freedom matches the number of
constraints\cite{Witten,Mouzarkel,wyart}, problematic; (2) forces at
the microscopic level are indeterminant due to friction and disorder;
(3) granular materials are athermal, so that conventional energy-based
statistical approaches are not appropriate; and (4) near isostaticity 
we expect fluctuations to be important, both within a
single realization of a system and from realization to realization.
New methods are needed to bridge the gap between force networks at
small length scales and continuum elasto-plastic theory at large
scales, and to capture the highly sensitive, fluctuating behavior of
granular systems near jamming.

We construct a model for stress fluctuations based on grain-scale
force and torque balance and positivity of contact forces rather than
energy conservation. We then calculate stress correlations and predict
differences for systems under isotropic compression versus shear
stress.  We also perform complementary numerical simulations and
experimental studies of jammed granular systems in 2D subject to
isotropic compression and pure shear. The stress correlation functions
from theory, simulation, and experiment are in qualitative and in some
cases quantitative agreement.  In particular, the theory predicts that
the form of the stress correlations depends on how the jammed states
were prepared.

{\bf Theoretical Framework:} Fluctuations are inherently related to the
number of microscopic states available under a given set of
macroscopic conditions.  In equilibrium thermodynamics, the
microcanonical entropy describes the nature of fluctuations and
response.  When we turn to granular systems, the identification of
states by energy is no longer useful, and a different criterion for
identifying states is needed.  
The approach that we pursue here
exploits a different conservation principal, based on force and torque
balance, which applies rigorously for granular
materials~\cite{silke-corey-bulbul, Blumenfeld07a, tighe-thiis}.  

The force-moment tensor of mechanically stable (MS) packings, $\hat
\Sigma = \int d^dr \hat \sigma(\mathbf{r})$, where $\hat \sigma
(\mathbf{r})$ denotes the local stress tensor, is an extensive
variable that is a topological
invariant~\cite{silke-corey-bulbul,silkethesis}.  In the force-moment
ensemble, $\hat \Sigma$ remains fixed barring system-spanning changes
in $\hat \sigma$.  Hence local fluctuations and response only involve
grain configurations with the same $\hat{\Sigma}$.  To construct a
theory for stress correlations, we adopt a coarse-grained approach, in
which jammed configurations are represented by a continuous
field~\cite{goldenfeld-book} and the entropy $S(\hat \Sigma)$ is
defined via an appropriate Ginzburg-Landau functional.  The theory
should be valid close to the jamming transition where grains
have negligible deformations, and stress fluctuations decouple from
volume fluctuations.  The decoupling is exact only in the limit of
infinitely rigid grains~\cite{makse-nature}.

In both two and three dimensions, a continuous field can be defined
that upholds force and torque
balance~\cite{blumenfeldball,silkethesis} of granular packings.  We
will focus on 2D systems, where a {\it scalar} field $\Psi$, the Airy
stress function~\cite{Landau}, is related to the local stress tensor by
\begin{equation}
\hat \sigma ({\bf r}) = \left [\begin{array}{cc}
\partial_{y}^{2}\Psi & -\partial_{x}\partial_{y}\Psi \\
-\partial_{x}\partial_{y}\Psi& \partial_{x}^{2} \Psi\\
\end{array} \right].
\label{eq:stress}
\end{equation}
We define $\Gamma = \mathrm{Tr} \hat \Sigma$ and $\tau =s_1 -s_2 =
\sqrt{\Gamma^2 - 4 ( \det \Sigma)^2}$, where $s_1> s_2$ are the
eigenvalues of $\hat \Sigma$.  Given a $\hat \Sigma$, $\Psi$ can be
expanded as $\Psi_0 + \psi$, where $\psi$ represents fluctuations
around $\Psi_0$.  The field $\Psi_0$ satisfies the biharmonic
equation, $\nabla^4 \Psi_0 = 0$~\cite{Landau}, and $\Sigma_{ij} = \int
d^d r(\delta_{ij}\nabla^{2} \Psi_0- \partial_{i} \partial_{j}
\Psi_0)$~\cite{silkethesis}.

The probability for fluctuations $\psi$ can be written as $P[\psi] =
Z^{-1}({\hat \Sigma}) e^{-{\sf L}[\Psi_0,\psi]} \equiv Z^{-1}({\hat
\Sigma})e^{-{\sf L}_{\hat \Sigma} [\psi]}$. The functional ${\sf
L}_{\hat \Sigma}[\psi]$ measures the contribution to the entropy from
all grain packings that have a coarse-grained representation
$\psi(\mathrm \bf r)$.  The partition function $Z({\hat \Sigma})
\equiv e^{S(\hat \Sigma)} = \int D\psi e^{-{\sf L}_{\hat
\Sigma}[\psi]}$ generates correlators of the field
$\psi$~\cite{goldenfeld-book}.  Because of gauge freedom, ${\sf
L}_{\hat \Sigma}[\psi]$ can only depend on second derivatives of $\psi$.
Independent second order scalars involving second derivatives of
$\psi$ can be constructed from the invariants of the local stress
tensor, $\mathrm{Tr}(\hat\sigma)$ and $\mathrm{det}(\hat\sigma)$.  The
coefficients of these terms and, therefore, the nature and strength of
fluctuations are controlled by the field $\Psi_0$ (or ${\hat \Sigma}$).
To lowest order in $\psi$, ${\sf L}_{\hat \Sigma}[\psi]$ resembles
the free energy for an elastic material in two
dimensions~\cite{Landau}:
\begin{eqnarray}      
{\sf L}_{\hat \Sigma}[\psi] = \int d^2r \, \Big\{ \alpha_1(\hat \Sigma)\, (\partial_x^2 \psi)^2 + \alpha_2(\hat \Sigma)\, (\partial_y^2 \psi)^2  \nonumber \\
+ \alpha_3(\hat \Sigma)\, (\partial_x \partial_y \psi)^2 + \alpha_4(\hat \Sigma)\, (\partial_x^2 \psi) (\partial_y^2 \psi) 
\Big\} ~.
\label{generalgl}
\end{eqnarray}      
The crucial differences between the description in Eq.~\ref{generalgl}
and traditional elasticity theory are (a) the stiffness constants are
determined by $\Psi_0$ or $\hat \Sigma$, and thus the theory is
inherently nonlinear, and (b) the origin of ${\sf L}_{\hat
\Sigma}[\psi]$ is entropic. The functional ${\sf L}_{\hat
\Sigma}[\psi]$ can be used to calculate averages and correlation
functions.  Below, we consider the behavior of pressure correlation
functions under isotropic compression and pure shear using
Eq.~(\ref{generalgl}).

\begin{figure}
\centering
\epsfig{file=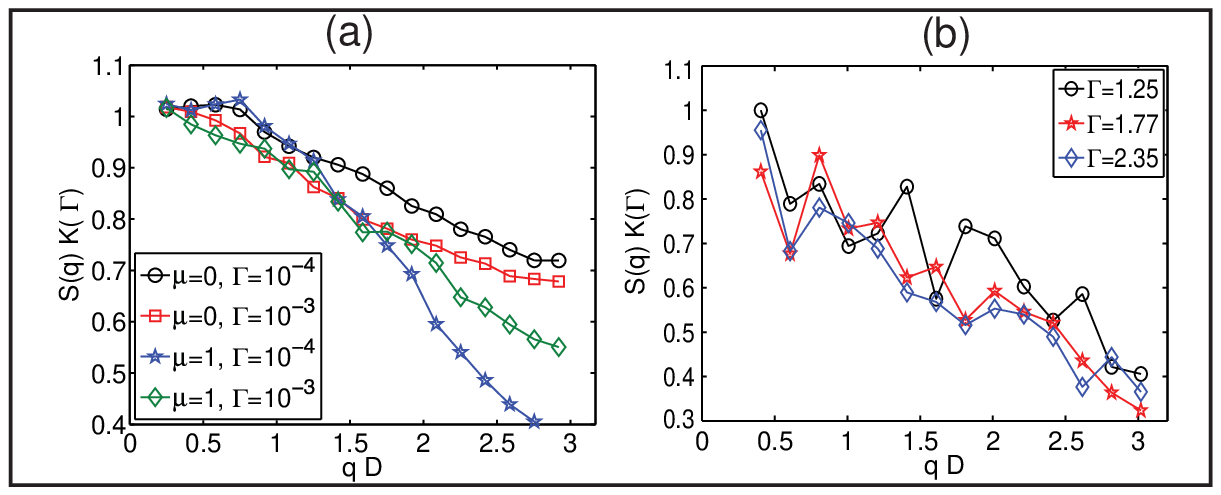,,width=0.9\linewidth,clip=} \\
\caption{\label{figure_qisotropic} Pressure correlations in Fourier
  space for pure compression. Angle-averaged $S(q) K(\Gamma)$ from (a)
  simulations at different $\Gamma$ (reduced units) and static
  friction coefficient $\mu$, (b) experiments at different $\Gamma$
  (in units of $\mathrm{N \cdot m}$), $\mu=0.7$ and $z$ ranging from
  $3.35$-$3.68$.  The experimental data have been scaled by $K(\Gamma)
  = ( z_{\rm iso}/2 + c(z - z_{\rm
  iso}^2))/\Gamma^2$\cite{silkethesis}, with $c=2.8$. Theory predicts
  that $S(q) K(\Gamma)$ is independent of $\Gamma$ for $q \ll 1/\xi$.}
\end{figure}

{\bf Isotropic Compression:} For isotropic compression with no
deviatoric stress, the stiffness constants only depend on $\Gamma$, and
${\sf L}_{\Gamma}[\psi]$ is isotropic:
\begin{eqnarray}
{\sf L}_\Gamma[\psi] & = &\int d^{2} r \lbrace {K(\Gamma) \over 2}\lbrace (\partial_{x}^{2}
\psi)^{2} +  (\partial_{y}^{2} \psi)^{2}\rbrace  \nonumber \\
&+ & \lambda (\Gamma) (\partial_{x}\partial_{y} \psi)^{2}  + (K(\Gamma) -\lambda(\Gamma)) \partial_{x}^{2}\psi
\partial_{y}^{2}\psi \rbrace,
\label{Freeiso}
\end{eqnarray}
with two stiffness constants $K$ and $\lambda$.  The positivity of
contact forces implies that both $\partial_{x}^{2} \Psi$ and
$\partial_{y}^{2} \Psi$ must be
non-negative\cite{silkethesis,silke-corey-bulbul}.  This is a
difficult constraint to impose exactly on the fluctuating field
$\psi$.  However, it can be enforced in a mean-field way by requiring
that $K(\Gamma) \geq 1/\Gamma^{2}$, which guarantees that the
amplitude of the long-wavelength fluctuations are such that the
positivity criterion is met\cite{silkethesis,Tighe-private}.  The
inequality constraint on $K(\Gamma)$ implies that different
preparation histories can lead to different fluctuations.  The maximum
entropy of jammed packings is achieved in protocols that meet the
equality, and we focus here on these marginal packings.  Note that the
positivity constraint does not impose any conditions on $\lambda$,
which is therefore taken to be independent of $\Gamma$.  Near jamming
when $\Gamma \rightarrow 0$, $K(\Gamma)$ becomes arbitrarily large and
the $\lambda$ terms can be ignored~\cite{footnote}.

The results for the correlations of the local pressure are best
visualized in Fourier space.  From Eq.~(\ref{Freeiso}), these
correlations are predicted to be isotropic:
\begin{eqnarray}
S({\bf q}) & =& <|\delta \Gamma({\bf q})|^{2}> = q^{4}
<|\psi({\bf q})|^{2}> \nonumber \\ &= & {K^{-1}(\Gamma)
\over {1 +\xi^{2} q^{2}}},
\label{pressureq}
\end{eqnarray}
where $\delta \Gamma = \nabla^2 \psi$, ${\bf q}$ is the wavevector,
and $\xi$ is a correlation length that describes the decay of
correlations at large $q$, and is defined by higher order terms not
included in Eq.~(\ref{Freeiso}).  In an experiment or simulation at
fixed $\Gamma/A$, there are many MS packings, and each is
characterized by a continuously varying field $\psi ({\bf r})$.  The
spatial correlations of stress, for a given $\Gamma$, are determined
by averaging over these configurations.  If the configurations are
sampled according to the theoretically predicted $P[\psi] \propto
e^{-{\sf L}_{\Gamma}[\psi]}$, the correlations measured in simulations
and experiments should be well described by the field-theoretic
predictions.  Since frictionless granular packings are isostatic near
jamming~\cite{ohern}, an exact calculation yields $K(\Gamma) = z_{\rm
iso}/(2 \Gamma^{2})$~\cite{silkethesis}, where the number of contacts
$z_{\rm iso}=4$ in 2D.  In contrast, frictional packings have $z_{\rm
iso}=3$ in 2D, but are often hypostatic~\cite{silbertgrest}.  We do
not have an exact result for $K(\Gamma)$ away from isostaticity,
however, a form that has been successful~\cite{silkethesis} is
$K(\Gamma) = ( z_{\rm iso}/2 + c(z - z_{\rm iso}^2))/\Gamma^2$, where
$c$ is a phenomenological constant.  We will use this form to compare
the predictions with results from frictional packings.

To test the $q$-space pressure fluctuations predicted in
Eq.~(\ref{pressureq}), we have numerically generated MS packings of
bidisperse disks ($N/2$ large and $N/2$ small particles with diameter
ratio $r=1.4$) both with and without friction near the jamming
transition using well-known packing-generation
algorithms~\cite{xuohern,maksepre}.  For frictionless grains, these
algorithms generate packings at the margin of stability~\cite{wyart}.
We studied system sizes ranging from $N=256$ to $4096$, systems with
square cells and periodic boundary conditions, pressures in the range
$\Gamma/A = 10^{-5}$ to $10^{-3}$ (in reduced units of the grain
stiffness), and static friction coefficients in the range $\mu=[0,1]$.
We have also carried out experiments using a biaxial apparatus that
has been described previously~\cite{RBTrush,trushprl,zhang}.  The biax
is a device that allows us to apply highly controlled deformations to
quasi-2D systems of photoelastic disks.  By using photoelastic disks,
it is possible to obtain all contacts and contact forces in the
system.  In this study, contact forces are calculated for $N=1228$
disks, with $N/5$ large and $4 N/5$ small disks with diameter
ratio $r=1.2$ and coefficient of static friction $\mu=0.7$.  The
experimental protocol generates packings farther from isostaticity
than those from the simulation protocol.

Results for pressure correlations in compressed systems are shown in
Fig.~\ref{figure_qisotropic}.  In general, we find that $S(q)$ decays
isotropically with ${\bf q}$.  In Fig.~\ref{figure_qisotropic}(a), we
plot the angle-averaged $S(q)$ from simulations, normalized by
$K(\Gamma)$ at $\Gamma/A = 10^{-3}$ and $10^{-4}$ as a function of
$qD$, where $D$ is the small particle diameter.  For both frictional
$\mu=1$ and frictionless $\mu=0$ grains, the results from the
simulations match Eq.~(\ref{pressureq}) at small $q$, with no fitting
parameters.  In Fig.~\ref{figure_qisotropic}(b) we plot the
angle-averaged $S(q) K(\Gamma)$ from experiments.  Both simulations
and experiments confirm the theoretically predicted scaling of $S(q)$
as $1/K(\Gamma)$ and increasing stiffness as the system is
decompressed.

{\bf Pure Shear:} In the presence of an imposed pure shear, the positivity
constraints on the stress lead to different conditions on the
stiffness constant $K$ in the $x$ and $y$ directions.  A dramatic
consequence is that the pressure correlations $S(q)$
become anisotropic even for infinitesimal shear, and the
correlations in real space become long-ranged.  To lowest order in
$\psi$ the entropy functional is
\begin{eqnarray}
{\sf L}_{\tau,\Gamma}[\psi] & = & \int d^{2} r \lbrace \lbrace {K(\Gamma + \tau) \over 2}
(\partial_{x}^{2}\psi)^{2} + {K(\Gamma - \tau) \over 2} (
\partial_{y}^{2}\psi)^{2} \rbrace \nonumber \\ &+ & \lambda
(\partial_{x}\partial_{y} \psi)^{2} ) + K^{\prime}(\Gamma, \tau)
\partial_{x}^{2}\psi \partial_{y}^{2}\psi \rbrace.
\label{Freeaniso}
\end{eqnarray}
Here, $x$ ($y$) is the principal axis of $\hat \Sigma$ with the smaller
(larger) eigenvalue.  In the case of pure shear, there are now two
distinct stiffness coefficients to ensure that both $\hat
\sigma_{xx}$ and $\hat \sigma_{yy}$ are positive.  In addition,
$K^{\prime}$ controls the entropy cost of fluctuations that contribute
to $\hat \sigma_{xx} \hat \sigma_{yy} $.  A more symmetric version can
be constructed by demanding that $K^{\prime} = \sqrt {(K(\Gamma +
  \tau) K(\Gamma - \tau) )}$, although we have no rigorous argument to
support this form.

\begin{figure}
\centering
\scalebox{0.73}{\includegraphics{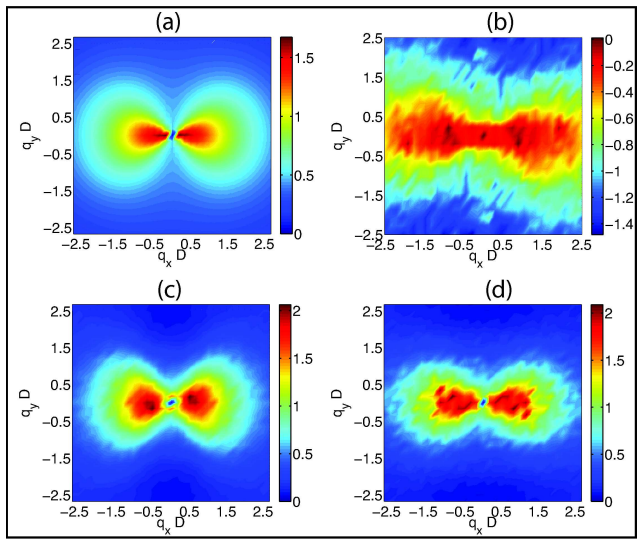}}
\caption{ \label{anisopressure-contours} Contours of $S(q) K(\Gamma)$
under pure shear: (a) theory, using $\tau/\Gamma = \xi/D = 0.3$; (b)
experiment, with intensity on a log-scale; and simulations of (c)
frictionless and (d) frictional ($\mu=1$) particles.  In both sets of
simulations, $\tau/\Gamma=0.3$.  In all plots, compression (dilation)
is along the vertical (horizontal) axis.}
\end{figure}

\begin{figure}
\centering
\epsfig{file=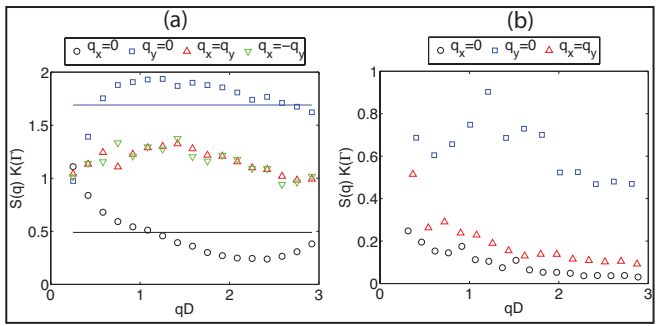,width=1\linewidth,clip=} \\
\caption{ \label{cuts} Cuts along axes specified in the legends for
$S(q) K(\Gamma)$ under pure shear.  (a) Simulation results with
$\mu=0$ and $\tau/\Gamma=0.3$. The solid lines are theoretical
predictions for $q_x=0$ and $q_y=0$.  Results for $\mu=1$ are
qualitatively similar.  (b) Experimentally measured $S(q) K(\Gamma)$
contours at $\tau/\Gamma=0.51$. The ratio of $S(q_y=0)/S(q_x =0)$ is
close to the theoretical prediction in (Eq. \ref{psiq-shear}), $1 +
4\frac{ \tau}{\Gamma} + O((\frac{ \tau}{\Gamma})^2) \simeq 3$.}
\end{figure}

The pressure correlations predicted from Eq.~(\ref{Freeaniso}) are:
\begin{eqnarray}
S(\!{\bf q}\!)& \!=\! & 
q^{4} ( K(\Gamma\!+\!\tau)q_{x}^{4}\!+\!K(\Gamma\!-\!
\tau)q_{y}^{4}\!+\!2 K^{\prime}(\Gamma) q_{x}^{2}q_{y}^{2} \nonumber \\
& + & K(\Gamma)\xi^{2}q^{6} )^{-1}. 
\label{psiq-shear}
\end{eqnarray}
The anisotropic, dipolar nature of this correlation function is
depicted in Fig.~\ref{anisopressure-contours}~(a).  To compare theory
with experiment, we create a sheared packing by first isotropically
compressing the system to a mechanically stable state at a density
slightly above jamming.  We then apply pure shear by expanding the
system in one direction while compressing in the other, keeping the
density constant.  The resultant pressure correlations are shown in
Fig~\ref{anisopressure-contours}~(b), and they match the expected
  form within the noise of the data.  To compare theory and
simulation, we generated MS packings of bidisperse disks with and
without friction over a range of stress ratios $\tau/\Gamma$.  To do
this, we compressed (dilated) the simulation cell in the $y$ ($x$)
direction by $\epsilon=\delta L/L$ over the range $\epsilon =
[10^{-5},10^{-3}]$.  Pressure correlations from simulations in
Fig.~\ref{anisopressure-contours}~(c) and (d) also show the dipolar
character of the pressure correlations. A key prediction of
Eq.~\ref{psiq-shear} is that $\lim _{{q \rightarrow 0}} S ({\bf q})$
depends on the direction of approach.  This feature is clearly
demonstrated in Fig.~\ref{cuts}, which shows the simulation and
experimental results for $S({\bf q})$ along different cuts in
$q-$space, along with the small-$q$ predictions from theory.  There is
good agreement between theory and simulation for $q_x=0$ and $q_y=0$,
where theoretical predictions exist.  Even though the theoretical
predictions make several simplifying assumptions such as $z-z_{\rm
  iso} \ll 1 $ (small $\Gamma$) and $\tau/\Gamma \ll 1$, we observe qualitative
agreement with experimental data.  In particular the pressure
correlations depend on the direction of approach to $q=0$ and they are
larger along $q_y=0$ than $q_x=0$.

The anisotropic nature of correlations in $q$-space imply
anisotropic decays in real space~\cite{silke-unpub} with a slower
decay along the direction of higher compression.  The entropic
formulation with the positivity constraint, therefore, provides an
explanation for the shear-induced anisotropy in pressure correlations
observed in experiments~\cite{RBTrush}.

{\bf Discussion:} We present a field theoretic approach, based on
entropy of packings, for describing stress fluctuations in granular
packings.  The theory enforces conditions of mechanical stability and
positivity of contact forces, and applies close to the jamming
transition, where grains have small deformations. From the theory, we
calculate pressure correlations and show that they depend sensitively
on the method used to generate the jammed state.  Under isotropic
compression, all correlations are isotropic and obey a simple scaling
relation as a function of compression.  For packings subjected to pure
shear, the correlations are anisotropic with a characteristic dipolar
feature in $q$-space.  The anisotropy is a consequence of the
positivity constraint, which causes $q$-space stress fluctuations to
be reduced along the compressive direction.  The present approach
provides a means of relating stress fluctuations to the
history of granular systems, which determines the force moment
tensor, and an explanation for the anisotropic behavior of stress
fluctuations.  The theoretical predictions for the pressure
correlation functions are confirmed, semiquantitatively, by
simulations of MS packings with and without friction and by
experiments on photoelastic disks.  This agreement is remarkable since
it validates the idea that the entropy of MS packings can be used to
determine the response of the this far-from-equilibrium system.

%Furthermore we find no evidence
%for a growing length scale near jamming~\cite{zorana}.  The Airy stress function has
%critical correlations for any compression and/or shear.  The deviation
%of isostaticity affects the stiffness constant but in a non singular,
%mild way: $K(\Gamma) \rightarrow K(\Gamma) (1 + c (z-z_{\rm iso})^{2})$.
%So, the $1/(z-z_{\rm iso}) $ length scale that shows up in the dynamical
%matrix simply renormalizes the stiffness, and the correction is small
%close to Point J \cite{wyart}.

Work supported by NSF-DMR0555431 (BC,SH), NSF-DMR0448838 (GL),
NSF-DMS0835742 (CO), NSF-DMR0555431 (JZ,TSM,RB), and YINQE (GL).  BC
acknowledges  discussions with Nick Read, and CS, BC, GL
acknowledge the Aspen Center for Physics and Lorentz Center, where
aspects of this work were performed.

%%% PRL


\begin{thebibliography}{100}
%\bibitem{nagelscience} C.-h. Liu {\it et al}, Science {\bf
%269}, 513 (1995).
\bibitem{nagelscience} C.-h. Liu, S. R. Nagel, D. A. Schecter, S. N.
Coppersmith, S. Majumdar, O. Narayan, and T. A. Witten, Science {\bf
269}, 513 (1995).
\bibitem{Dinsmore} J. Zhou, S. Long, Q. Wang and A. D. Dinsmore,
Science {\bf 312}, 1631 (2006).
\bibitem{nagel} D. M. Mueth, H. M. Jaeger, and S. R. Nagel,
Phys. Rev. E {\bf 57}, 3164 (1998).
\bibitem{dantu} P. Dantu, Ann. Ponts Chaussees {\bf 4}, 144 (1967).
\bibitem{jgeng} J. Geng, D. Howell, E. Longhi and R. P. Behringer,
G. Reydellet, L. Vanel, E. Cl\'{e}ment, and S. Luding,
Phys. Rev. Lett. {\bf 87}, 035506 (2001).
\bibitem{RBTrush} T. S. Majmudar and R. P. Behringer, Nature {\bf
435}, 1079 (2005).
\bibitem{Landau} L. D. Landau and E. M. Lifshitz, Theory of
Elasticity, (Butterworth-Heinemann, London, 1986).
\bibitem{Bouchaud} M. Otto, J.-P. Bouchaud, P. Claudin, and
J. E. S. Socolar, Phys. Rev. E {\bf 67}, 031302 (2003).
\bibitem{Cates} M. E. Cates, J. P. Wittmer, J.-P. Bouchaud and
P. Claudin, Phys. Rev. Lett. {\bf 81}, 1841 (1998).
\bibitem{Blumenfeld} R. Blumenfeld, Phys. Rev. Lett. {\bf 93}, 108301 (2004).
\bibitem{GoldGold} I. Goldhirsch and C. Goldenberg, Eur. Phys. J. E {\bf 9}, 245-251 (2002).
\bibitem{Witten}
A. V. Tkachenko and T. A. Witten, Phys. Rev. E {\bf 60}, 687 (1999).
\bibitem{Mouzarkel}
C. F. Moukarzel, Phys. Rev. Lett. {\bf 81}, 1634 (1998).
\bibitem{wyart}
M. Wyart, S. R. Nagel, and T. A. Witten, Europhys. Lett. {\bf 72}, 486 (2005).
\bibitem{silke-corey-bulbul} S. Henkes, C. S. O'Hern and B. Chakraborty,
Phys. Rev. Lett. {\bf 99}, 038002 (2007).
\bibitem{Blumenfeld07a}
R. Blumenfeld in {\it Lecture Notes in Complex Systems Vol 8: Granular and
Complex Materials}, edited by T. Aste, A. Tordessilas and T. D. Matteo (2007).
\bibitem{tighe-thiis}
B. P. Tighe, A. R. T. van Eerd and T. J. H. Vlugt,
Phys. Rev. Lett. {\bf 100}, 238001 (2008).
\bibitem{makse-nature} C. Song, P. Wang and H. A. Makse, Nature {\bf
453}, 629 (2008).
\bibitem{blumenfeldball} R. C. Ball and R. Blumenfeld,
Phys. Rev. Lett. {\bf 88}, 115505 (2002).
\bibitem{silkethesis} S. Henkes, Ph.D. dissertation, (2008); S. Henkes and B. Chakraborty, to appear in Phys. Rev. E. (2009).
\bibitem{goldenfeld-book}
N. Goldenfeld, ``Lectures on Phase Transitions and the Renormalization Group'',  (Addison-Wesley, New York, 1992).
\bibitem{Tighe-private} B. Tighe, private communication.
\bibitem{footnote}
Higher-order derivatives of $\psi$ will likely enter if $K \rightarrow \infty$
and suppress stress fluctuations.
\bibitem{ohern} C. S. O'Hern, L. E. Silbert, A. J. Liu and
S. R. Nagel, Phys. Rev. E {\bf 68}, 011306 (2003).
\bibitem{silbertgrest} L. E. Silbert, D. Ertas, G. S. Grest, T. C. Halsey and D. Levine, Phys. Rev. E
{\bf 65}, 031304 (2002).
\bibitem{xuohern} N. Xu, J. Blawzdziewicz and C. S. O'Hern,
Phys. Rev. E {\bf 71}, 061306 (2005).
\bibitem{maksepre} H. P. Zhang and H. A. Makse, Phys. Rev. E 
{\bf 72}, 011301 (2005).
\bibitem{trushprl} T. S. Majmudar, M. Sperl, S. Luding and R. P. Behringer, Phys. Rev. Lett. {\bf 98}, 058001 (2007).
\bibitem{zhang} J. Zhang, T. S. Majmudar, A. Tordesillas, and R. P. Behringer,
preprint (2009).
\bibitem{silke-unpub} S. Henkes, unpublished (2009).

\end{thebibliography}
\end{document}